# Inflammatory Bowel Disease Biomarkers of Human Gut Microbiota Selected via Ensemble Feature Selection Methods

Hilal Hacilar[1], O.Ufuk Nalbantoglu[2], Oya Aran[3], Burcu Bakir-Gungor[1]

**Abstract**

The tremendous boost in the next generation sequencing and in the "omics" technologies makes it possible to characterize human gut microbiome —the collective genomes of the microbial community that reside in our gastrointestinal tract—. While some of these microorganisms are considered as essential regulators of our immune system, some others can cause several diseases such as Inflammatory Bowel Diseases (IBD), diabetes, and cancer. IBD, comprising Crohn's disease and ulcerative colitis, is a gut related disorder where the deviations from the "healthy" gut microbiome are considered to be associated with IBD. Although some existing studies attempt to unveal the composition and functional capacity of the gut microbiome in relation to IBD diseases, a comprehensive picture of the gut microbiome in IBD patients is far from being complete. Due to the complexity of metagenomic studies, the applications of the state-of-the-art machine learning techniques became popular to address a wide range of questions in the field of metagenomic data analysis. In this regard, using IBD associated metagenomics dataset, this study utilizes both supervised and unsupervised machine learning algorithms, i) to generate a classification model that aids IBD diagnosis, ii) to discover IBD-associated biomarkers, iii) to find subgroups of IBD patients using k-means and hierarchical clustering. To deal with the high dimensionality of features, we applied robust feature selection algorithms such as Conditional Mutual Information Maximization (CMIM), Fast Correlation Based Filter (FCBF), min redundancy max relevance (mRMR) and Extreme Gradient Boosting (XGBoost). In our experiments with 10-fold cross validation, XGBoost had a considerable effect in terms of minimizing the microbiota used for the diagnosis of IBD and thus reducing the cost and time. We observed that compared to the single classifiers, ensemble methods such as kNN+logitboost resulted in better performance measures for the classification of IBD.

[1] Dept. of Comp. Eng., Abdullah Gul University, Kayseri, Turkey

[2] Dept. of Comp. Eng, Genome & Stem Cell Cnt, Erciyes Uni., Kayseri, Turkey

3 College of Computer Studies, De La Salle Univ., Manila, Philippines

## 1. Introduction

Human gut microbiota is a complex community of microorganisms including trillions of bacteria that populate in our gastrointestinal tract. While some of these microorganisms are considered as essential regulators of our immune system, some others can infect human body and cause diseases such as IBD, diabetes, obesity, cancer, autoimmune and neurodegenerative disorders. In this respect, deciphering the function and composition of our gut microbiome —the collective genomes of the microbial community that reside in human gut—is crucial.

In recent decades, the rapid advances in next generation sequencing (NGS) technologies enabled to generate millions to billions of reads in a single run. Metagenomic NGS approaches, which permit to analyze the entire genomic content of a sample and provide taxonomic and functional profiles of microbial communities; accelerated the discovery of the human gut microbiome. Since our gut microbiome is modulated via human-microbiome symbiosis, the metagenomic analysis of gut microbiome provides novel insights regarding the effect of human gut microbiota on human physiology and diseases [1-5].

IBD, comprising Crohn's disease (CD) and ulcerative colitis (UC), is a gut related disorder that affects the intestinal tract. Deviations from the "healthy" gut microbiome that occur due to environmental effects, dietary and genetic mutations are considered to be associated with IBD. Hence, the metagenomic analysis of human gut microbiome helps to illuminate IBD development mechanisms. Since the aetiology of IBD is not fully understood and symptoms are complex, the design of new tools that make use of the available human gut metagenome data is essential for the diagnosis of IBD. In this respect, machine learning (ML) is well suited to obtain a diagnostic model using IBD-associated metagenomics dataset.

Mosotto *et.al.* [6] suggested to use machine learning algorithms to diagnose CD and UC patients among a group of paediatric inflammatory bowel disease patients. Using endoscopic and histological data, they achieved 82.7% CD/UC discrimination accuracy. Pasolli *et. al.* [7] attempted to classify the cases (patients) and the controls (healthy samples) using the metagenomic-associated datasets of Cirrhosis, Colorectal Cancer, IBD, Obesity and Type 2 diabetes. They tested the performances of the support vector machines (SVM), random forest (RF) classifiers and also evaluated Lasso and elastic net regularized multiple logistic regression. Due to the low number of CD patients, they combined CD and UC patients together and their predictive model achieved the area under the curve (AUC) of 0.914 for discrimination of IBD patients using the species abundance information as a feature vector. Yazdani *et. al.* [8] investigated the major changes in microbiome protein family abundances between IBD patients and healthy subjects. Wei *et. al.* [9] proposed a risk prediction scheme for IBD. Using genome-wide association study data obtained from 17,000 CD cases, 13,000 UC cases, and 22,000 controls, they obtained AUCs of 0.86 and 0.83 for CD and UC, respectively. Wingfield *et. al.* [10] presented a method for the stratification of IBD presence, and IBD subtype from a bacterial census of the intestinal microbiome. Via using a hybrid classifier of multi layer perceptron (MLP) and SVM, they obtained AUCs of 0.70 and 0.74 for CD and UC, respectively.

In our previous work [11], for the diagnosis of IBD, we analyzed IBD-associated metagenomics dataset using state-of-art machine learning algorithms, ensemble methods and shrinkage methods including ridge regression, Lasso [12] and Elastic Net [13]. Our best models achieved AUC of 0.919, 87.7 % accuracy and F1-measure of 83.7 %.

In this study, firstly, we attempt to identify IBD-associated biomarkers via utilizing robust feature selection algorithms such as Conditional Mutual Information Maximization (CMIM), Fast Correlation Based Filter (FCBF), Min redundancy Max relevance (mRMR) and Extreme Gradient Boosting (XGBoost). Using state-of-the art machine learning algortihms and ensemble classification methods such as SVM, Decision tree, RF, Adaboost, and Logitboost, we assess the performance of our reduced metagenomics dataset. We evaluate our models systematically and extensively using several metrics. Secondly, we aim to find subgroups of IBD patients via applying k-means and hierarchical clustering on IBD-associated metagenomics dataset. Since the symptoms and the treatments of IBD are complex, the precise identification of IBD subgroups could provide a valuable insight for discovering individualized therapy targets and will pave the way towards personalized medicine applications. Additionally, we perform Principal Compenent Analysis (PCA) to obtain the underlying structure of IBD metagenomics data. In summary, this study utilizes both supervised and unsupervised machine learning algorithms, *i)* to generate a classification model that aids IBD diagnosis, *ii)* to investigate potential pathobionts of IBD, and *iii)* to find out subgroups of IBD patients.

The rest of this paper is organized as follows. Section 2 presents the machine learning algorithms and feature selection methods that we used to obtain a diagnostic model of IBD, to identify IBD biomarkers of human gut microbiota, and to discover the subgroups of IBD patients. Section 3 highlights our findings and provides an extensive evaluation of our method. Section 4 presents a discussion of our findings. Section 5 concludes the paper and summarizes avenues for further research.

## 2. Materials and Methods

In this study, we aim to develop a classification model to aid IBD diagnosis and to discover IBD-associated biomarkers using metagenomics data. In this respect, the raw microbiome DNA sequencing data of 148 IBD patients and 234 control samples were fetched from MetaHit project [14] and categorized into disease states based on the associated metadata. Using MetaPhlAn2 taxonomic classification tool, each DNA sequence was assigned to its microbial species of origin (taxa). Consequently, the microbial diversity (i.e. which microorganisms exist in what relative abundance) of the gut microbiome for each sample was revealed. As shown in Figure 1, our IBD-associated metagenomics dataset included the sequence reads from 1455 taxa for 382 samples. This dataset is used to develop, train, test and validate a machine-learning model to diagnose IBD.

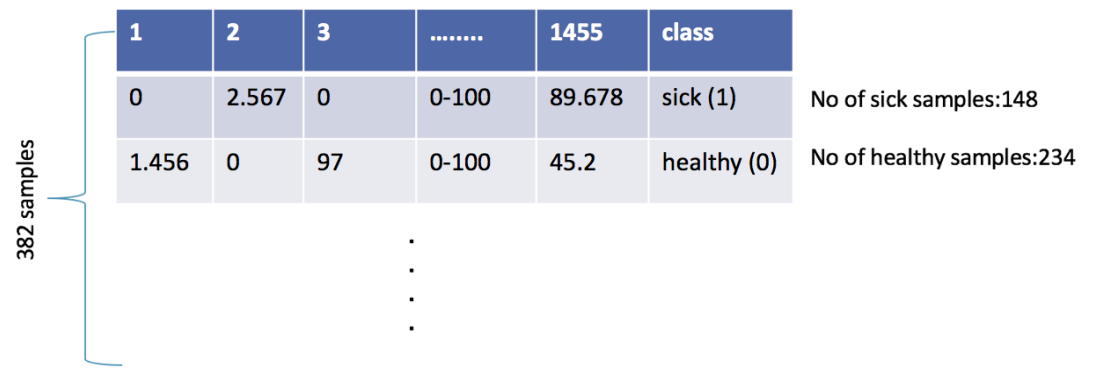

**Figure 1.** Illustration of the metagenomics dataset

Our methodology can be summarized in the following steps: i) *feature selection* to identify the most informative IBD-associated biomarkers, ii) *model construction* to classify IBD patients and control samples, followed by the assessment of the generated models using extensive evaluation metrics, iii) *unsupervised learning* to detect subgroups of IBD patients and control samples. Figure 2 shows the details of these three steps.

### 2.1 Feature Selection

In metagenomics studies, the number of predictors (number of taxa) is much more than the number of observations. In this respect, feature selection methodologies such as mRMR [15], Lasso [7], Elastic Net [13] and iterative sure select algorithm [15] have been extensively applied to reduce the number of taxa (species). In this work, we proposed to apply Conditional Mutual Information Maximization (CMIM), Fast Correlation Based Filter (FCBF) and Min Redundancy Max Relevance (mRMR), Extreme Gradient Boosting (XGBoost) feature selection algorithms on metagenomics dataset, which have not been used in this domain before. CMIM first ranks the features according to their conditional entropy and mutual information with the class to predict; and then selects the feature if it carries additional information. Similarly, FCBF ranks features based on their mutual information with the class to predict; and then removes the features whose mutual information is less than a predefined threshold. mRMR selects the features that have the most correlation with a class to predict (relevance) and the less correlation between themselves (redundancy). In XGBoost feature selection, the more an attribute is used to make key decisions with decision trees, the higher relative importance it gets. Via applying a predefined threshold, one can select top ranked features in CMIM, FCBF, and XGBoost.

### 2.2 Model Construction

In order to discriminate IBD samples from controls, we construct a range of machine learning models using different classification algorithms such as RF, Desicion Tree, Logitboost, Adaboost, SVM and stacking ensemble classifiers. Python scikit-learn library [16] is used for the implementation of these algorithms.

Prediction performances of our models were evaluated by using Accuracy, F1 Score, and AUC measures. Accuracy is a widely used performance evaluation metric and a reliable measure for the balanced datasets. Since we have an uneven class distribution in our dataset, to evaluate the performance of our models we also utilize other metrics such as F1 score and the area under the ROC curve (AUC). F1-score, which denotes the harmonic mean of presicion and recall, is a better performance metric, when someone seeks a balance between precision and recall, and when there is an uneven class distribution (large number of actual negatives). AUC is commonly used as a summary measure of diagnostic accuracy. In real life examples, there is an overlap between the test results of positive and negative examples. AUC shows how the recall vs. precision relationship changes as we change the threshold or the cut-off value for identifying a positive.

### 2.3 Unsupervised Learning

The relationship between the samples and the species is important to understand the disease development mechanisms, to detect subgroups of IBD patients and controls. In order to visualize and analyze this relationship, we performed three unsupervised methods: k-means clustering, PCA, and hierarchical clustering using euclidian distance.

## 3. Experiments

### 3.1 Classification

Our IBD-associated metagenomics dataset included the sequence reads from 1455 taxa for 382 samples. Firstly, we removed the irrelevant and redundant features (species) using the feature selection strategies including FCBF, CMIM, mRMR and XGBoost. Applied on IBD-associated metagenomics dataset, Figure 3 illustrates the numbers of selected features using different feature selection algorithms. While 56 of 1455 features are selected by at least one of the four different feature selection algorithms, 10 features are selected by all of the tested feature selection algorithms.

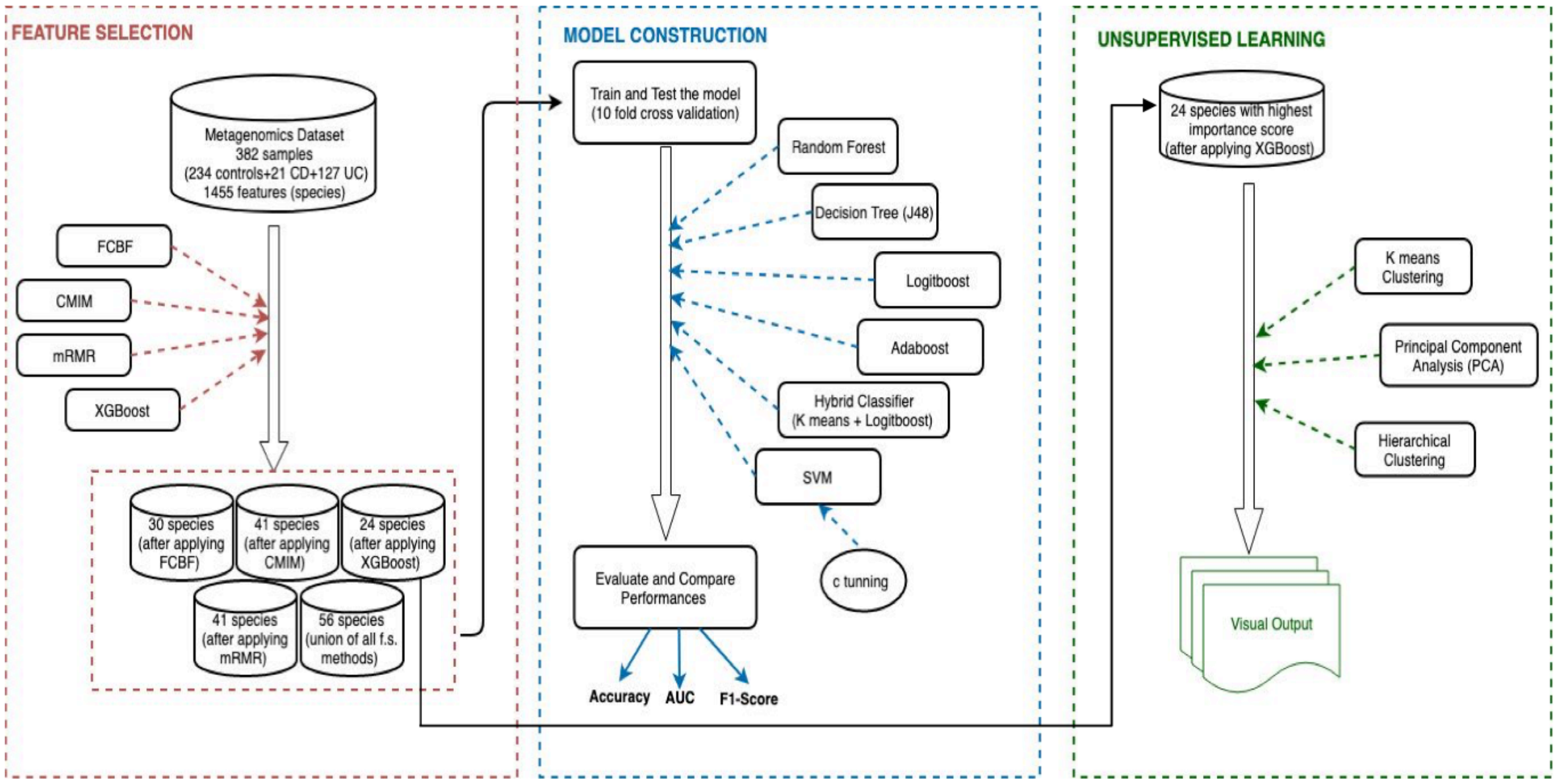

**Figure 2.** Schematic representation of our methodology. Feature selection, model construction and unsupervised learning processes are shown in red, blue and green, respectively.

Secondly, we optimize the parameters “c” and “gamma” for SVM, “number of tree” for Desicion tree, Random Forest, Logitboost and Adaboost. Following the identification of optimal parameters and features, we validate and test our models using 10-fold cross validation. Lastly, we assessed the performance of our models using the features that are selected by each feature selection algorithm independently, and also using the union of the features, which are chosen by at least one of the four different feature selection algorithms.

Our models are extensively evaluated using several evaluation metrics, as summarized in Figure 4. We observed that ensemble methods such as kNN (k nearest neighbor) + Logitboost resulted in better performance measures for the classification of IBD patients vs control samples. For the diagnosis of IBD patients, significant results, i.e., 0,947 AUC, 0,872 F1-score and 90,05% accuracy were obtained by using 24 features (selected using XGBoost) and stacking with k means and logitboost classifier.

Figure 4 highlights the improvement in terms of performance measures once the feature selection strategies are applied (Fig. 4a-d, 4f). For all tested classifiers, XGBoost feature selection algortihm resulted in higher performance values with only 24 features (Fig 4d). The feature selection methods, especially XGBoost, is found to have a considerable effect in terms of minimizing the microbiota used for the diagnosis of IBD and thus reducing the cost and time. Also, the union of the selected features (56 features) had a considerable performance improvement (Fig. 4f).

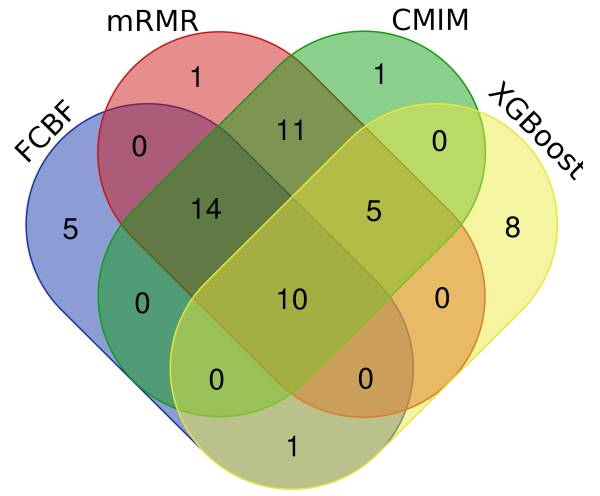

**Figure 3.** Numbers of selected features (species) using different feature selection algorithms

Figure 5 illustrates the comparision of the performances of all tested classification algorithms on IBD-associated metagenomics dataset in terms of feature selection methods. As shown in Figure 5, the application of the selected feature selection algorithms increased the performances of the classification algorithms. The union of the selected features (56 features) shows the best performance in terms of F-measure. In terms of ROC area, the classification algorithms including RF, Logitboost, Adaboost and ensemble classifiers resulted in higher performances (>0.9) on our IBD associated metagenomics dataset.

(a)

| Classification Algorithm | TP Rate | FP Rate | Precision | Recall | F-Measure | ROC Area | Accuracy |
|---|---|---|---|---|---|---|---|
| Random Forest | 0,797 | 0,098 | 0,837 | 0,797 | 0,817 | 0,934 | 86,1257 |
| Decision Tree (J48) | 0,682 | 0,179 | 0,706 | 0,682 | 0,694 | 0,762 | 76,7016 |
| LogitBoost | 0,845 | 0,098 | 0,845 | 0,845 | **0,845** | **0,939** | **87,9581** |
| AdaBoost | 0,797 | 0,094 | 0,094 | 0,797 | 0,819 | 0,931 | 86,3874 |
| Stacking (kmeans+opt.SVM) | 0,757 | 0,128 | 0,789 | 0,757 | 0,772 | 0,846 | 82,7225 |
| Stacking (kmeans+Logitboost) | 0,838 | 0,103 | 0,838 | 0,838 | 0,838 | 0,905 | 87,4346 |
| SVM (param Opt.+normalized polykernel) | 0,77 | 0,132 | 0,786 | 0,77 | 0,778 | 0,819 | 82,9843 |

(b)

| Classification Algorithm | TP Rate | FP Rate | Precision | Recall | F-Measure | ROC Area | Accuracy |
|---|---|---|---|---|---|---|---|
| Random Forest | 0,811 | 0,085 | 0,857 | 0,811 | 0,833 | 0,936 | 87,4346 |
| Decision Tree (J48) | 0,669 | 0,167 | 0,717 | 0,669 | 0,692 | 0,771 | 76,9634 |
| LogitBoost | 0,845 | 0,085 | 0,862 | 0,845 | **0,853** | **0,941** | **88,7435** |
| AdaBoost | 0,811 | 0,085 | 0,857 | 0,811 | 0,833 | 0,933 | 87,4346 |
| Stacking (kmeans+opt.SVM) | 0,743 | 0,094 | 0,833 | 0,743 | 0,786 | 0,867 | 84,2932 |
| Stacking (kmeans+Logitboost) | 0,838 | 0,098 | 0,844 | 0,838 | 0,841 | 0,917 | 87,6963 |
| SVM (param Opt.+normalized polykernel) | 0,764 | 0,09 | 0,843 | 0,764 | 0,801 | 0,837 | 85,3403 |

(c)

| Classification Algorithm | TP Rate | FP Rate | Precision | Recall | F-Measure | ROC Area | Accuracy |
|---|---|---|---|---|---|---|---|
| Random Forest | 0,811 | 0,085 | 0,857 | 0,811 | 0,833 | 0,937 | 87,4346 |
| Decision Tree (J48) | 0,669 | 0,171 | 0,712 | 0,669 | 0,69 | 0,772 | 76,7016 |
| LogitBoost | 0,824 | 0,085 | 0,859 | 0,824 | 0,841 | **0,941** | 87,9581 |
| AdaBoost | 0,811 | 0,081 | 0,863 | 0,811 | 0,836 | 0,935 | 87,6963 |
| Stacking (kmeans+Opt.SVM) | 0,757 | 0,09 | 0,842 | 0,757 | 0,797 | 0,87 | 85,0785 |
| Stacking (kmeans+Logitboost) | 0,851 | 0,077 | 0,875 | 0,851 | **0,863** | 0,917 | **89,5288** |
| SVM (param Opt.+normalized polykernel) | 0,723 | 0,09 | 0,836 | 0,723 | 0,775 | 0,817 | 83,7696 |

(d)

| Classification Algorithm | TP Rate | FP Rate | Precision | Recall | F-Measure | ROC Area | Accuracy |
|---|---|---|---|---|---|---|---|
| Random Forest | 0,831 | 0,077 | 0,872 | 0,831 | 0,851 | 0,947 | 88,7435 |
| Decision Tree (J48) | 0,716 | 0,167 | 0,731 | 0,716 | 0,724 | 0,74 | 78,7958 |
| LogitBoost | 0,858 | 0,077 | 0,876 | 0,858 | 0,867 | **0,956** | 89,7906 |
| AdaBoost | 0,818 | 0,081 | 0,864 | 0,818 | 0,84 | 0,942 | 87,9581 |
| Stacking (kmeans+opt.SVM) | 0,709 | 0,132 | 0,772 | 0,709 | 0,739 | 0,885 | 80,6283 |
| Stacking (kmeans+Logitboost) | 0,872 | 0,081 | 0,872 | 0,872 | **0,872** | 0,947 | **90,0524** |
| SVM (param Opt.+normalized polykernel) | 0,709 | 0,124 | 0,784 | 0,709 | 0,745 | 0,793 | 81,1518 |

(e)

| Classification Algorithm | TP Rate | FP Rate | Precision | Recall | F-Measure | ROC Area | Accuracy |
|---|---|---|---|---|---|---|---|
| Random Forest | 0,676 | 0,077 | 0,847 | 0,676 | 0,752 | 0,915 | 82,7225 |
| Decision Tree (J48) | 0,696 | 0,167 | 0,725 | 0,696 | 0,71 | 0,743 | 78,0105 |
| LogitBoost | 0,73 | 0,068 | 0,871 | 0,73 | 0,794 | **0,934** | 85,3403 |
| AdaBoost | 0,669 | 0,073 | 0,853 | 0,669 | 0,75 | 0,917 | 82,7225 |
| Stacking (kmeans+opt.SVM) | 0,682 | 0,145 | 0,748 | 0,682 | 0,714 | 0,832 | 78,7958 |
| Stacking (kmeans+Logitboost) | 0,77 | 0,09 | 0,844 | 0,77 | **0,806** | 0,917 | **85,6021** |
| SVM (param Opt.+normalized polykernel) | 0,689 | 0,141 | 0,756 | 0,689 | 0,721 | 0,774 | 79,3194 |

(f)

| Classification Algorithm | TP Rate | FP Rate | Precision | Recall | F-Measure | ROC Area | Accuracy |
|---|---|---|---|---|---|---|---|
| Random Forest | 0,824 | 0,081 | 0,865 | 0,824 | 0,844 | 0,943 | 88,2199 |
| Decision Tree (J48) | 0,682 | 0,171 | 0,716 | 0,682 | 0,699 | 0,764 | 77,2251 |
| LogitBoost | 0,851 | 0,068 | 0,887 | 0,851 | 0,869 | **0,951** | 90,0524 |
| AdaBoost | 0,818 | 0,09 | 0,852 | 0,818 | 0,834 | 0,941 | 87,4346 |
| Stacking(kmeans+opt.SVM) | 0,662 | 0,073 | 0,852 | 0,662 | 0,745 | 0,824 | 82,4607 |
| Stacking(kmeans+Logitboost) | 0,878 | 0,06 | 0,903 | 0,878 | **0,89** | 0,933 | **91,623** |
| SVM (param Opt.+normalized polykernel) | 0,75 | 0,12 | 0,799 | 0,75 | 0,774 | 0,815 | 82,9843 |

**Figure 4.** Performance evaluations of different classifiers on IBD metagenomics dataset, using 10-fold cross validation with parameter optimization, after applying (a) FCBF, (b) CMIM, (c) mRMR, (d) XGBoost feature selection techniques, (e) without feature selection, (f) union of FCBF, CMIM, mRMR and XGBoost.

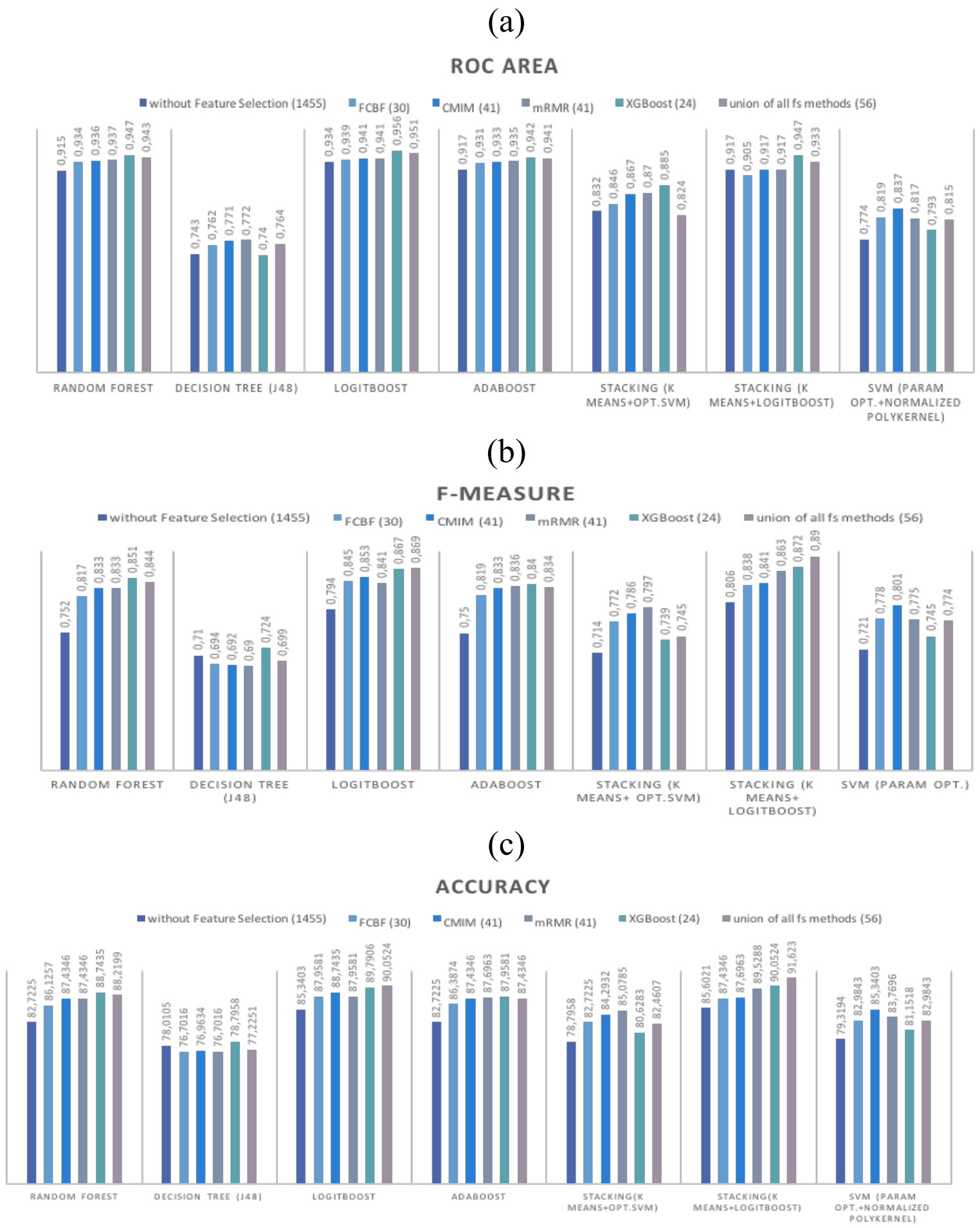

**Figure 5**. Comparisons of the performances of different classifiers and different feature selection methods, in terms of **(a)** Area under ROC, **(b)** F-Measure, and **(c)** Accuracy

### 3.2 Feature importances and their correlations

The identification of critical species, the ones having a key role in IBD disease development, can constitute new targets for the development of probiotics to correct microbiota aberrations [17-20]. To this end, we analyzed the XGBoost importance scores of the species and found that some of the identified species are related with IBD development mechanisms. Figure 6 shows the feature importances of the 24 features (species) that are selected via XGBoost. In Figure 6a, while the Y axis corresponds to feature ids of the features, X axis corresponds to relative importances which are calculated using information gain. Figure 6b lists the high scoring 24 features with their species names.

In order to analyze the pairwise correlations of these selected species, we use Spearman's rank correlation coefficient and illustrate these relations using a heatmap in Figure 6c. This analysis resulted in only two moderate positive correlations between the two species namely *Eubacterium Hallii - Coprococcus Comes* and *Eubacterium Hallii – Dorea formicigenerans*.

### 3.3 Grouping control and IBD samples via K means clustering algorithm

We investigate that if we subgrouped the samples according to their health situations, whether these subgroups have a direct relationship with some species. In this respect, we use K means clustering algortihm to subgroup samples.

K means is an unsupervised learning algorithm that groups samples by minimizing the error inside the clusters and maximizing the distance between the clusters. We select the number of clusters using elbow method. In this method, the point where the decline in the error slows down indicates the optimum number of clusters. As shown in Figure 7, using euclidian distance, optimum 12 subgroups among controls and 7 subgroups among IBD samples were discovered. Figure 8 illustrates the identified control and IBD subgroups and the presence of the species in each of these subgroups.

It can be concluded from Figure 8 that even though the samples were divided into subgroups, a single species has no direct effect on the development of IBD.

Nevertheless, there are a few observations that we can make: i) *Bifidobacterium-bifidum* (shown in orange) is mainly observed in IBD subgroups (ibd 01, 06, 07) and not in the control subgroups; ii) *Bifidobacterium adolecentis* is observed in ibd06 and ibd07 and in none of the control groups except cont04; iii) *Bacteroides stercoris* (shown in gray) is present in five of the ibd groups (ibd 01-05), where there is no significanct presence of *Bifidobacterium adolecentis*. Consecutively, *Bacteroides stercoris* (shown in gray) is not present in ibd06 and ibd07, where there is high presence of *Bifidobacterium adolecentis*.

### 3.4 Discrimination of data via Principal Compenent Analysis (PCA)

In order to analyze whether the metagenomic data can be divided into two clusters representing control and IBD samples, the first two principal compenents of data were obtained using PCA as an unsupervised learning approach.

PCA is used for reducing the dimension of the feature space in such a way that the principal compenents in the new feature space are orthogonal to each other. The PCA results in Figure 9 show that better seperation is observed between control and IBD classes when PCA is performed with 24 top scoring features (Figure 9b, 9d). Also, it is important to point out that this result is achieved only with 24 species instead of 1455 species (Figure 9a, 9c). However, the new feature space, reduced via PCA, does not have a significant contribution in terms of classification performance. It is important to note that since PCA maps the data into a new feature space, the original feature information is lost during this process. Thus, PCA analysis does not allow for any biomarker discovery, as the species information is no longer represented in the new mapped feature space.

**(a)**

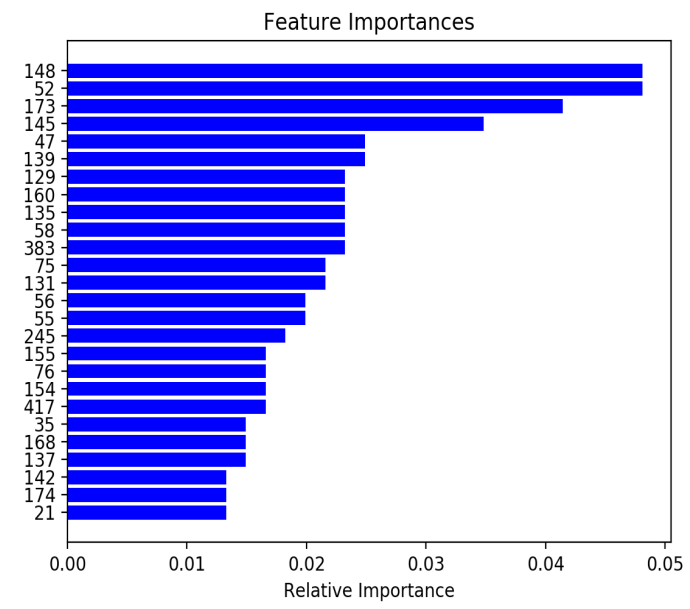

**(b)**

**148:** *Lachnospiraceae bacterium GCF 000218445*
**52:** *Bacteroides xylanisolvens*
**173:** *Ruminococcus bromii*
**145**: *Dorea formicigenerans*
**47:** *Bacteroides stercoris*
**139:** *Butyrivibrio crossotus*
**129:** *Eubacterium hallii*
**160:** *Roseburia inulinivorans*
**135:** *Blautia hydrogenotrophica*
**58:** *Parabacteroides distasonis*
**383:** *Clostridiales Family XIII Incertae Sedis unclassified*
**75:***Alistipes finegoldii*
**131:***Eubacterium rectale*
**56**:*Odoribacter splanchnicus*
**55**:*Coprobacter fastidiosus*
**245:***Akkermansia muciniphila*
**155:***Lachnospiraceae bacterium GCF 000242155*
**76:***Alistipes indistinctus*
**154:***Lachnospiraceae bacterium GCF 000185525*
**417:***Peptostreptococcus stomatis*
**35:***Bacteroides dorei*
**168:***Faecalibacterium prausnitzii*
**137:***Ruminococcus obeum*
**142:***Coprococcus comes*

**(c)**

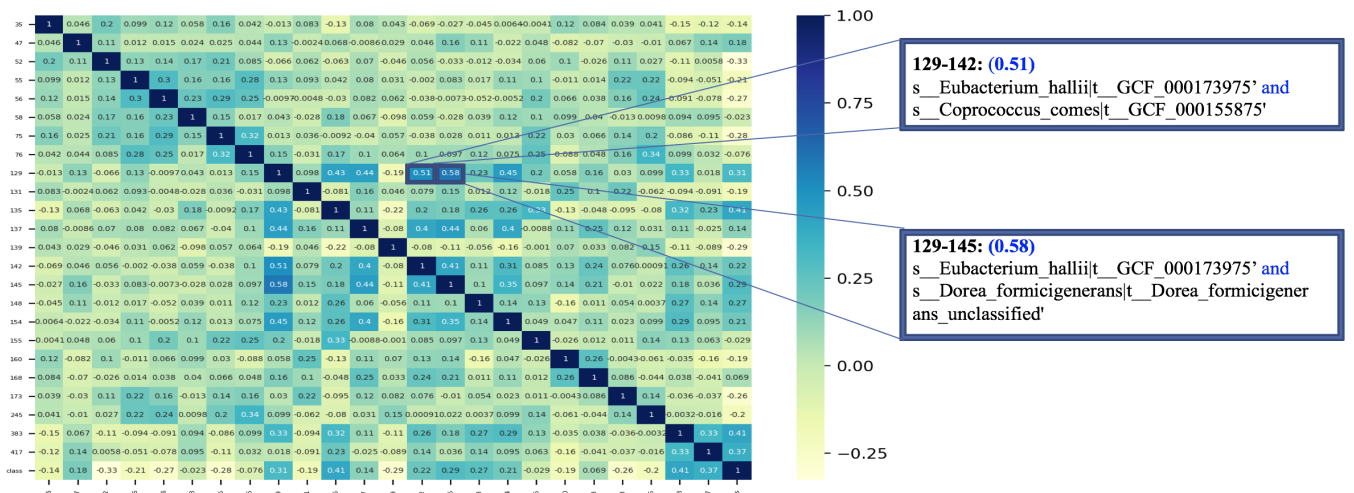

**Figure 6. (a)** The features that have high relative importance scores in XGBoost. While the Y axis corresponds to feature ids of these features, X axis corresponds to relative importance values of corresponding features. The sum of relative importance of all 1455 features is equals to 1. **(b)** List of the species that have top 24 importance scores for IBD. **(c)** Pairwise Spearman rank correlations of top 24 important species.

**(a)** **(b)**

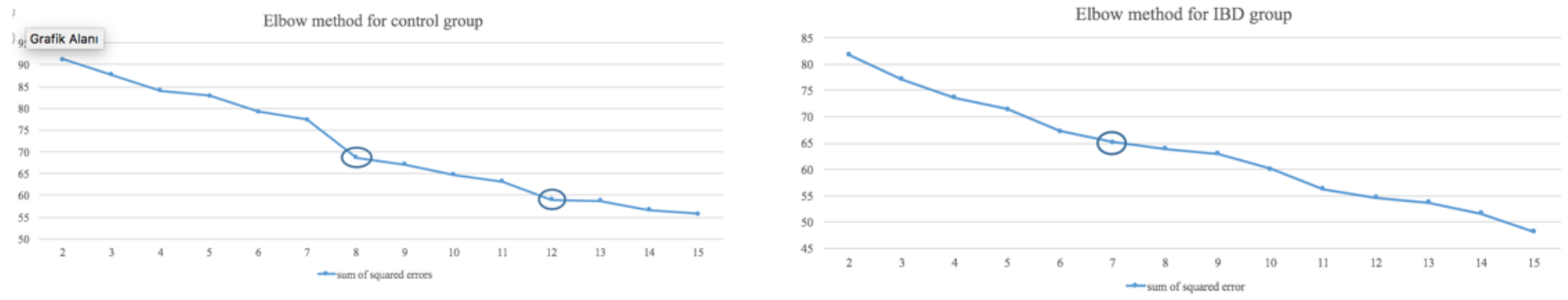

**Figure 7.** Selection of the optimum number of subgroups for **(a)** controls and **(b)** IBD patients.

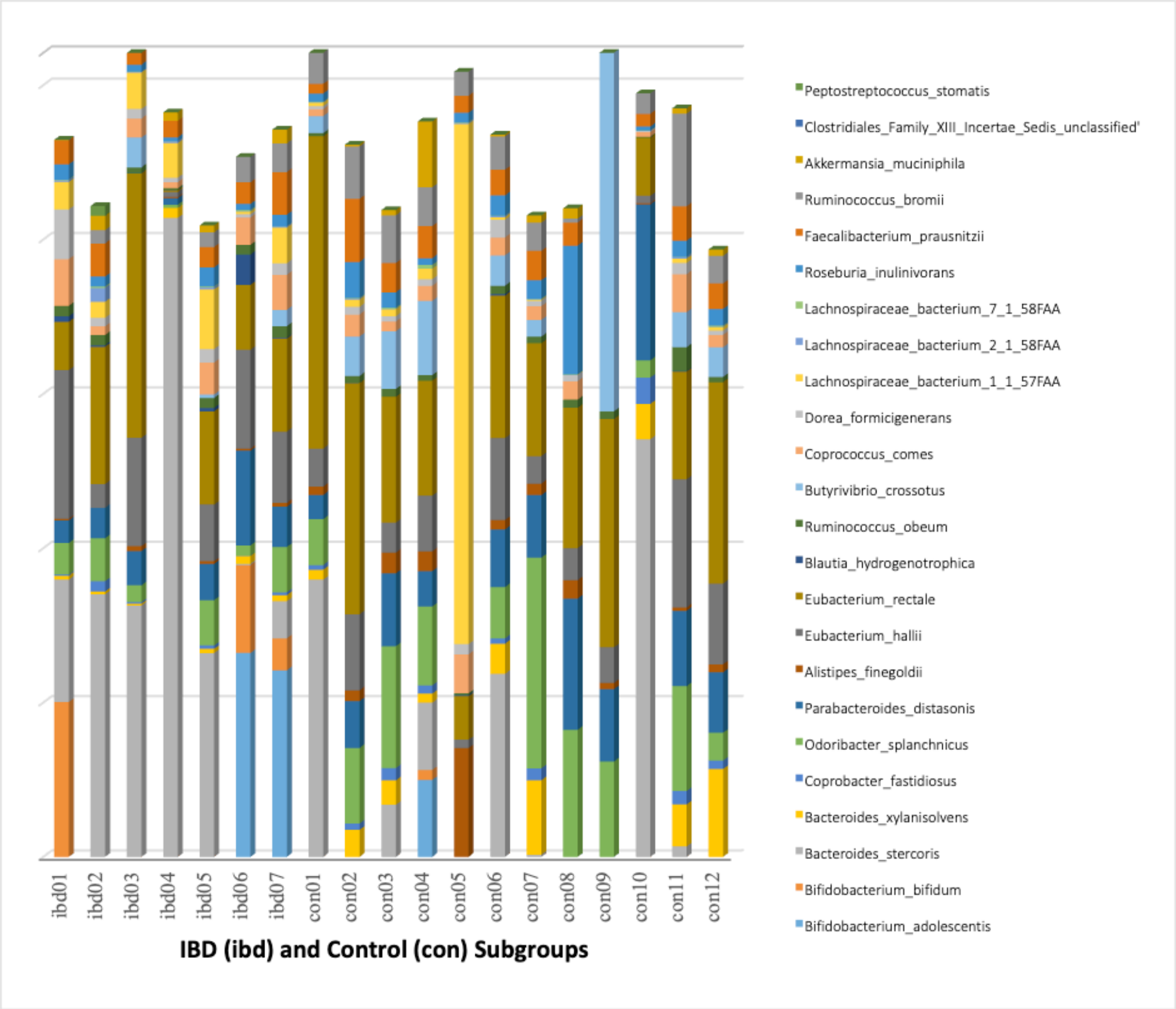

**Figure 8**. Illustration of control and IBD subgroups (best viewed in color).

### 3.5 Hierarchical Clustering of reduced IBD dataset

In order to better visualize the relationship between the samples and the species, we also performed hierarchical clustering using euclidian distance and Ward variance minimization algorithm as the linkage method. The heatmap in Figure 10 is obtained using all samples and top scoring 24 species. Colors represent raw based z-scores of all samples. While the black color indicates read counts just around the mean, the lighter colors denote the read counts of 1 to 4 standart deviations above the mean. The first column specifies class labels, i.e., IBD patients and healthy samples are shown in red and blue, respectively. The areas that are restricted with red boxes could suggest the relative abundance of the corresponding species in these subgroups. For example, the red boxes in 5$^{th}$ and 6$^{th}$ columns indicate excessive levels of *Lachnospiraceae bacterium* and *Bacteroides stercoris* in the corresponding IBD subgroups.

## 4. Discussion

In this study, we compared the performances of machine learning methods including SVM, RF Adaboost, Logitboost, Decision tree and some ensemble methods on a dataset, which is obtained from MetaHit project containing species as features of the human gut microbiota of IBD patients and controls. To deal with the high dimensionality of features, we apply some feature selection methods including FCBF, CMIM, mRMR and XGBoost. We analyzed the XGBoost importance scores of the species and the top scoring species are found to be related with IBD development mechanisms. Among the species that we identified as potential IBD biomarkers (listed in Figure 6b), *Bacteroides xylanisolvens* [21] and *Eubacterium hallii* [22] are considered as candidate next-generation probiotics, promoting gut health. Other selected features of bacterial taxa, *Lachnospiraceae, Parabacteroides, Blautia, Butyrivibrio, Dorea, Ruminococcus*, and *Roseburia* are previously identified as potential biomarkers of IBD [23]. However, discovering these markers and potential therapeutic agents either requires laborious wet-lab processes or they can be only discovered at genus level. Our study enabled a shotgun discovery of multiple biomarkers via applying a careful feature selection.

Our proposed feature selection procedures achieve unprecedented performance using a feature subset of smaller cardinality, compared to the current state-of-the-art. Using a similar IBD-associated metagenomics dataset which is obtained from MetaHIT Project, Passolli *et. al.* [7] achieved 0.809% accuracy, 0.81 recall, 0.89 AUC, 0.78 precision and 0.78 F1 score, when they reduced the number of species into 441 using gini index. In this work, unprecedented results such as 91.623% accuracy, 0.933 AUC, 0.89 F1 score, 0.903 precision and 0.878 recall were obtained by using the union for selected features (a total of 56 features) and stacking with K means and logitboost classifier. Using only the XGBoost feature selection algorithm (24 features) with the same classifier achieves 90.05% accuracy, 0.947 AUC, 0.872 F1 score, 0.872 Precision, and 0.872 recall. These selected species could be suggested as potential IBD-biomarkers of human gut microbiata. Hence, this study leads to a framework for precise biomarker discovery, which enables targeting minimal number of diagnostic/theurepathic markers with large effect sizes. Such metagenomic drug discovery tools would imply a potential for narrowing the spectrum intervention, which can be feasible with the current pharmaceutical technology.

## 5. Conclusions

Gut microbiota can affect the host immune system and metabolism, which are central to program many aspects of host activities. Metagenomic analysis of human microbiome reveals significant phenotypical signals, such as disease, as microbiome is modulated via human-microbiome symbiosis. Since the accuracy of diagnosis in IBD is key to prompt an effective treatment, there is an utmost need to develop a novel classification technique that can expedite IBD diagnosis. This study utilizes several supervised and unsupervised machine learning algorithms on an IBD associated metagenomics data to aid diagnostic accuracy, to investigate potential pathobionts of IBD, and to find out which subset of microbiota is more informative than other taxa applying some of the state-of-the art feature selection methods. Overall, this paper provides a blueprint for the use of advanced feature selection and machine learning techniques on disease-associated metagenomics data.

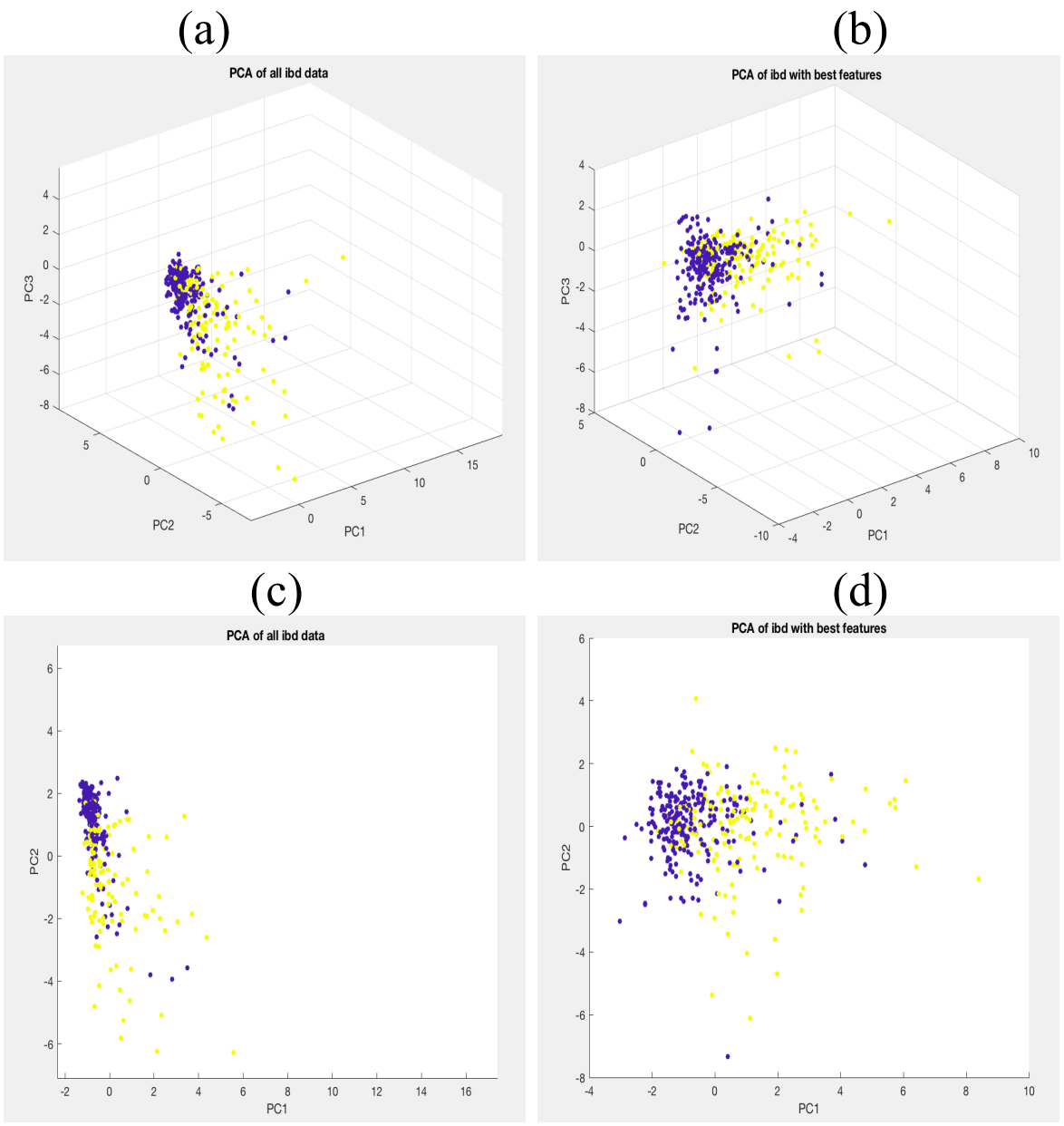

**Figure 9.** Principal component analysis of **(a, c)** all IBD-associated metagenomics dataset, **(b, d)** reduced dataset that includes features for top scoring 24 species.

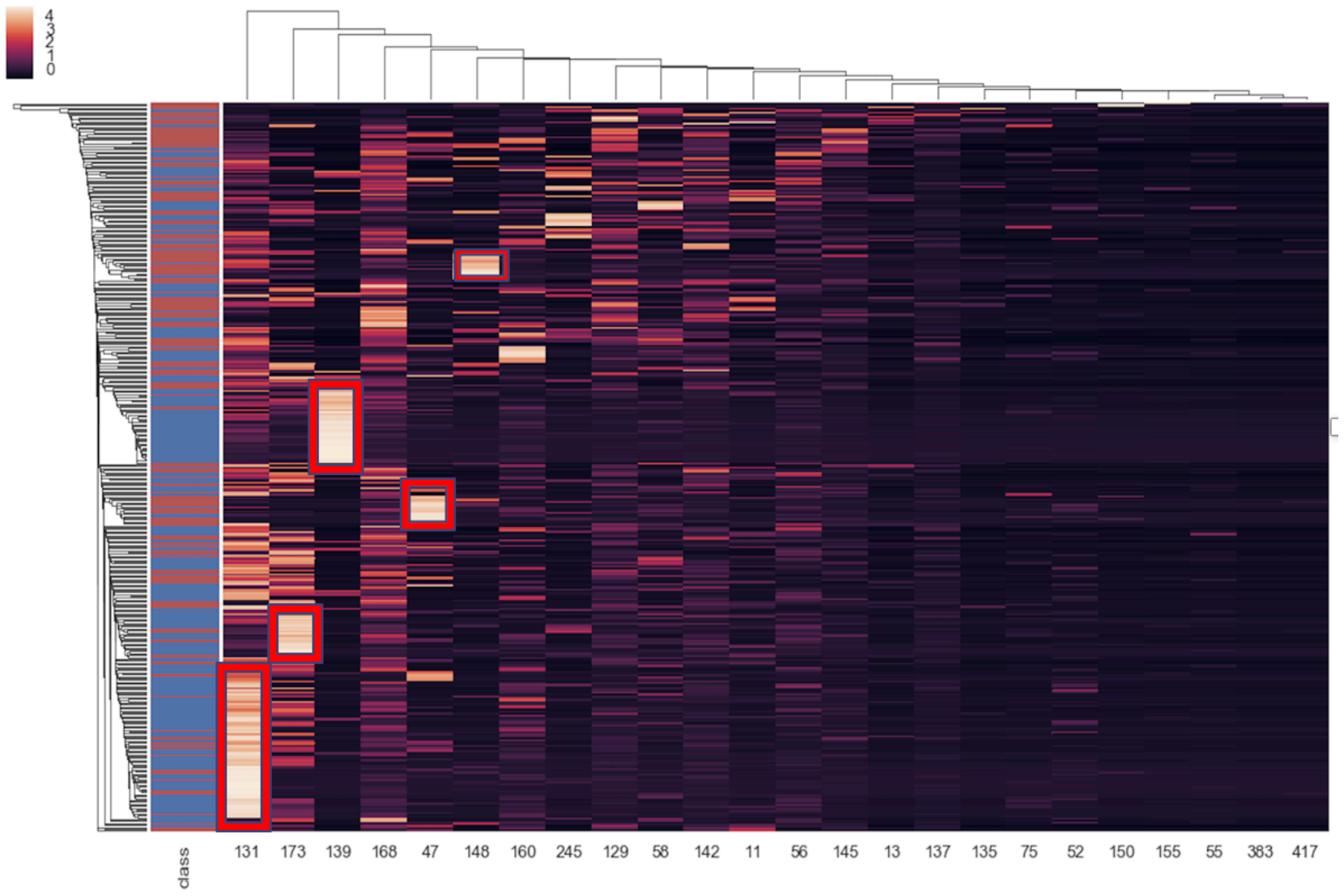

**Figure 10.** Z-score normalization (raw based) and Euclidian distance based hierarchical clustering. Ward variance minimization algorithm was used as a linkage method. The side bar on the left handside (named class) indicates the referred diagnosis: IBD patients and healthy samples are shown in red and blue, respectively. The dendograms are plotted on the top and left side of the figure.